\documentstyle{cupconf}
\input epsf

\ifoldfss
\else
  \ifnfssone
    \newmathalphabet{\mathit}
      \addtoversion{normal}{\mathit}{cmr}{m}{it}
      \addtoversion{bold}{\mathit}{cmr}{bx}{it}
    \newmathalphabet{\mathcal}
      \addtoversion{normal}{\mathcal}{cmsy}{m}{n}
    \else
    \ifnfsstwo
    \fi
  \fi
\fi

\def\spose#1{\hbox to 0pt{#1\hss}}
\def\lta{\mathrel{\spose{\lower 3pt\hbox{$\sim$}}
    \raise 2.0pt\hbox{$<$}}}
\def\gta{\mathrel{\spose{\lower 3pt\hbox{$\sim$}}
    \raise 2.0pt\hbox{$>$}}}

\title[Inside-Out Bulge Formation]{Inside-Out Bulge Formation \\
and the Origin of the Hubble Sequence}

\author[F. C. van den Bosch]{Frank C. van den Bosch}

\affiliation{Department of Astronomy, University of Washington,
Box 351580, Seattle, WA 98195, USA}

\begin{document}

\ifnfssone
\else
  \ifnfsstwo
  \else
    \ifoldfss
      \let\mathcal\cal
      \let\mathrm\rm
      \let\mathsf\sf
    \fi
  \fi
\fi

\maketitle

\begin{abstract}
  Galactic disks are thought to originate from the cooling of baryonic
  material inside virialized dark halos.   In order for these disks to
  have  scalelengths  comparable to observed  galaxies,  the  specific
  angular momentum   of the baryons    has to  be  largely conserved.  
  Because  of  the  spread  in  angular   momenta  of  dark   halos, a
  significant fraction of disks are expected to be  too small for them
  to be   stable, even if no angular   momentum is lost.  Here   it is
  suggested that a self-regulating  mechanism is at work, transforming
  part of the baryonic material into a bulge,  such that the remainder
  of   the  baryons can settle   in   a stable  disk  component.  This
  inside-out bulge formation   scenario  is coupled  to   the Fall  \&
  Efstathiou theory of disk formation to search for the parameters and
  physical   processes  that  determine  the  disk-to-bulge ratio, and
  therefore explain to  a  large  extent   the origin of   the  Hubble
  sequence. The  Tully-Fisher  relation   is used to     normalize the
  fraction  of baryons that  forms    the galaxy, and  two   different
  scenarios are investigated   for   how this  baryonic material    is
  accumulated  in  the center  of the  dark   halo. This simple galaxy
  formation  scenario can account for both  spirals and S0s, but fails
  to incorporate more bulge dominated systems.
\end{abstract}

\firstsection % if your document starts with a section,
              % remove some space above using this command.

\section{Introduction}

Despite considerable progress in our understanding of the formation of
galaxies, the  origin of the Hubble sequence  remains a major unsolved
problem.    The     main morphological    parameter    that   sets the
classification of galaxies in the  Hubble diagram is the disk-to-bulge
ratio  ($D/B$).  Understanding the   origin of the Hubble  sequence is
thus intimately related to understanding  the parameters and processes
that   determine the  ratio  between the  masses of  disk  and bulge.  
Especially, we need to understand  whether this ratio is imprinted  in
the  initial conditions   (`nature')    or whether  it  results   from
environmental processes  such   as  mergers  and impulsive  collisions
(`nurture').

Here I suggest a simple inside-out formation scenario for the bulge (a
`nature'-variant) and investigate the differences in properties of the
proto-galaxies that result  in different disk-to-bulge  ratios. A more
detailed discussion  on the background and   ingredients of the models
can be found in van den Bosch (1998; hereafter vdB98).

\section{The formation scenario}

In the standard picture of galaxy formation, galaxies form through the
hierarchical  clustering of dark matter  and subsequent cooling of the
baryonic matter  in the dark halo  cores.  Coupled with the  notion of
angular   momentum   gain  by     tidal torques   induced   by  nearby
proto-galaxies, this theory provides the  background  for a model  for
the formation of galactic disks.  In  this model, the angular momentum
of  the baryons  is  assumed to be  conserved  causing the  baryons to
settle in a rapidly  rotating disk  (e.g.,  Fall \& Efstathiou 1980).  
The turn-around, virialization, and subsequent cooling of the baryonic
matter  of  a   proto-galaxy  is an  inside-out    process.  First the
innermost shells  virialize and heat    its baryonic material  to  the
virial temperature. The cooling time of this  dense, inner material is
very short,  whereas its specific angular momentum  is relatively low. 
If the cooling time of the gas is shorter than the dynamical time, the
gas will condense  in clumps that  form stars, and this  clumpiness is
likely to result in  a   bulge.  Even   if the low-angular    momentum
material accumulates   in a disk,  the self-gravity  of  such a small,
compact disk  makes  it violently unstable, and  transforms  it into a
bar.  Bars are efficient  in transporting gas  inwards, and can  cause
vertical  heating by means of  a collective bending instability.  Both
these processes lead  ultimately to the  dissolution of the bar; first
the bar takes a hotter, triaxial shape,  but is later transformed in a
spheroidal bulge component.  There is thus  a natural tendency for the
inner, low  angular    momentum baryonic material  to   form   a bulge
component rather than a disk.   Because of the ongoing  virialization,
subsequent shells   of material cool and   try to settle   into a disk
structure at  a radius determined by  their angular  momentum.  If the
resulting  disk is unstable, part  of the material is transformed into
bulge   material.    This   process   of   disk-bulge   formation   is
self-regulating in that the bulge grows  until it is massive enough to
sustain  the remaining gas  in the form  of a stable  disk.  I explore
this inside-out bulge formation scenario, by incorporating it into the
standard Fall \& Efstathiou theory for disk formation.

The   {\it ansatz} for the  models  are the properties  of dark halos,
which are assumed to follow the universal density profiles proposed by
Navarro, Frenk \&  White   (1997),  and whose halo   spin  parameters,
$\lambda$, follow a  log-normal distribution in concordance  with both
numerical and analytical  studies.     I assume that only   a  certain
fraction,  $\epsilon_{\rm gf}$, of the  available  baryons in a  given
halo ultimately  settles   in the   disk-bulge  system.  Two   extreme
scenarios for this galaxy formation (in)efficiency are considered.  In
the first  scenario, which I   call the `cooling'-scenario,  only  the
inner fraction $\epsilon_{\rm  gf}$  of the  baryonic mass  is able to
cool  and form the  disk-bulge  system: the outer  parts of  the halo,
where the density is lowest, but which contain the largest fraction of
the  total  angular  momentum, never   gets  to cool.   In  the second
scenario,  referred  to  hereafter   as the   `feedback'-scenario, the
processes related to feedback and star  formation are assumed to yield
equal probabilities, $\epsilon_{\rm gf}$, for  each baryon in the dark
halo, independent of its initial radius  or specific angular momentum,
to  ultimately  end up   in the   disk-bulge  system.   The values  of
$\epsilon_{\rm gf}$  are normalized by  fitting the model disks to the
zero-point of the observed  Tully-Fisher relation. Recent observations
of   high redshift   spirals  suggest that     the zero-point of   the
Tully-Fisher relation does not evolve with redshift. This implies that
the  galaxy formation efficiency,  $\epsilon_{\rm  gf}$, was higher at
higher  redshifts (see   vdb98 for  details).    Disks are modeled  as
exponentials with a  scalelength proportional  to $\lambda$ times  the
virial radius of the halo  (as in the  disk-formation scenario of Fall
\& Efstathiou).   The bulge mass is  determined  by requiring that the
disk  is stable.  Since  the amount  of self-gravity  of  the disk  is
directly related  to the amount  of  angular momentum  of the gas, the
disk-to-bulge ratio in this scenario is mainly  determined by the spin
parameter of the dark halo out of which the galaxy forms.

\section{Clues to the formation of bulge-disk systems}

%
% Figure 1: Results. OCDM
%
\begin{figure}
  \epsfysize=7.9cm
  \centerline{\epsfbox{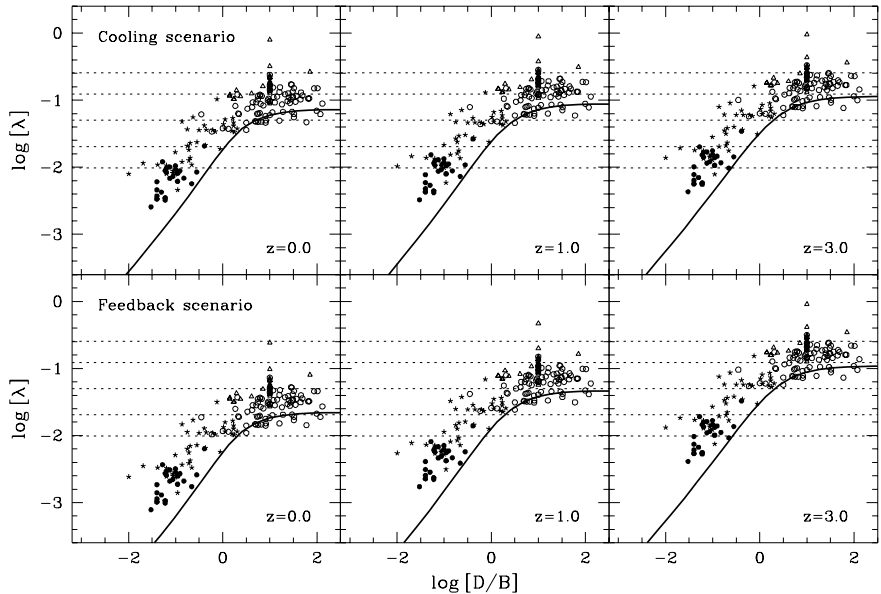}}
  \caption{Results for a OCDM cosmology with $\Omega_0 = 0.3$. Plotted are the
    logarithm of the spin parameter versus the logarithm of the
    disk-to-bulge ratio.  Solid circles correspond to disky
    ellipticals, stars to S0s, open circles to HSB spirals, and
    triangles to LSB spirals.  The thick solid line is the stability
    margin; halos below this line result in unstable disks.  As can be
    seen, real disks avoid this region, but stay relatively close to
    the stability margin, in agreement with the self-regulating bulge
    formation scenario proposed here.  The dashed curves correspond to
    the 1, 10, 50, 90, and 99 percent levels of the cumulative
    distribution of the spin parameter. Upper panels correspond to the
    cooling scenario, and lower panels to the feedback scenario.
    Panels on the left correspond to $z = 0$, middle panels to $z =
    1$, and panels on the right to $z = 3$.}
  \label{f1}
\end{figure}

Constraints on the formation scenario envisioned above can be obtained
from a comparison of these disk-bulge-halo models  with real galaxies. 
From the  literature I compiled   a list   of  $\sim 200$   disk-bulge
systems,  including  a wide variety  of   galaxies: both high  and low
surface brightness spirals (HSB  and LSB respectively), S0,  and disky
ellipticals (see vdB98 for details).  After choosing a cosmology and a
formation redshift, $z$, I calculate,  for each galaxy in this sample,
the spin   parameter  $\lambda$  of the   dark  halo which,   for  the
assumptions underlying  the formation  scenario proposed here,  yields
the   observed   disk properties  (scale-length    and central surface
brightness).   We thus use   the formation scenario  to  link the {\it
  disk}  properties  to those of   the dark halo,   and  use the known
statistical properties of dark halos to discriminate between different
cosmogonies.

The  main results are  shown  in Figure~1,  where  I plot the inferred
values of $\lambda$  versus the observed  disk-to-bulge  ratio for the
galaxies in  the sample.  The   dotted lines outline the  distribution
function of   halo spin  parameters of  dark   halos; it can  thus  be
inferred what the   predicted distribution of  disk-to-bulge ratios is
for galaxies that form  at a  given  formation redshift.   Results are
presented for an open cold  dark matter (OCDM)  model with $\Omega_0 =
0.3$ and no  cosmological  constant ($\Omega_{\Lambda}  = 0$).   These
results are virtually  independent of the value of $\Omega_{\Lambda}$,
but depend strongly on $\Omega_0$, which sets the baryon mass fraction
of the Universe.  Throughout,  a universal baryon density of $\Omega_b
= 0.0125  \, h^{-2}$ is  assumed,  in  agreement with  nucleosynthesis
constraints.   The  inferred spin  parameters   are larger  for higher
values of the assumed formation redshifts.  This owes to the fact that
halos  that virialize  at higher   redshifts  are denser.   Since  the
scalelength of the disk is  proportional to $\lambda$ times the virial
radius   of the halo, higher   formation  redshifts imply larger  spin
parameters in order  to yield the  observed disk  scalelength.  In the
cooling scenario, the probability that a certain  halo yields a system
with a  large  disk-to-bulge ratio (e.g.,  a spiral)  is rather small. 
This is due to the fact that in this scenario most of the high angular
momentum material never gets to cool to become  part of the disk.  The
large observed fraction of spirals in the field, renders this scenario
improbable.    For the feedback  cosmogony,  however, a more promising
scenario unfolds: At high redshifts ($z \gta 1$) the majority of halos
yields systems   with  relatively small   disks  (e.g., S0s),  whereas
systems that form  more    recently are more   disk-dominated   (e.g.,
spirals). This difference owes to two effects.  First of all, halos at
higher  redshifts are denser, and  secondly, the redshift independence
of the  Tully-Fisher  relation implies  that  $\epsilon_{\rm gf}$  was
higher at higher redshifts.  Coupled to the notion that proto-galaxies
that collapse at high redshifts  are preferentially found in overdense
regions such  as clusters, this scenario  thus  automatically yields a
morphology-density relation, in which S0s  are predominantly formed in
clusters of galaxies, whereas spirals are more confined to the field.

\section{Conclusions}

\begin{itemize}

\item Inside-out  bulge formation is a natural  by-product of the Fall
  \& Efstathiou theory for disk formation.

\item  Disk-bulge systems do  not  have bulges that are  significantly
  more massive than required by stability of the disk component.  This
  suggests a coupling between the formation of  disk and bulge, and is
  consistent with the   self-regulating,  inside-out bulge   formation
  scenario proposed here.

\item A comparison  of the angular momenta  of dark  halos and spirals
  suggests that   the baryonic material that builds   the disk can not
  loose a significant fraction   of its angular momentum.   This rules
  against the `cooling scenario' envisioned here, in which most of the
  angular momentum remains in the baryonic material in the outer parts
  of the halo that never gets to cool.

\item If we live in a low-density  Universe ($\Omega_0 \lta 0.3$), the
  only efficient way to make spiral  galaxies is by assuring that only
  a relatively small fraction of  the available  baryons make it  into
  the  galaxy,  and furthermore that the   probability  that a certain
  baryon becomes   a constituent    of the  final  galaxy has    to be
  independent of its specific  angular  momentum, as described by  the
  `feedback scenario'.

\item If more extended observations confirm that the zero-point of the
  Tully-Fisher relation is  independent  of redshift, it  implies that
  the galaxy formation efficiency,  $\epsilon_{\rm gf}$, was higher at
  earlier times.  Coupled with the  notion that density  perturbations
  that   collapse early are   preferentially   found in high   density
  environments  such as   clusters, the scenario  presented  here then
  automatically predicts a   morphology-density relation in  which S0s
  are most likely to be found in clusters.

\item A reasonable variation   in formation redshift and halo  angular
  momentum can yield approximately one order of magnitude variation in
  disk-to-bulge ratio, and the simple formation scenario proposed here
  can  account for  both spirals and   S0s. However, disky ellipticals
  have too large bulges and too small disks to be incorporated in this
  scenario.   Apparently, their formation   and/or evolution has  seen
  some processes that caused the baryons to loose a significant amount
  of their angular momentum.   Merging and impulsive collisions (e.g.,
  galaxy harassment)  are  likely to  play a   major role  for   these
  systems.

\end{itemize}

\medskip
 
It  thus seems that {\it  both  `nature' and `nurture' are accountable
  for the formation of spheroids}, and that  the Hubble sequence has a
hybrid origin.

\smallskip
  
%
% Acknowledgements
%
\begin{acknowledgments}
  Support for this work was provided by NASA through Hubble Fellowship
  grant \# HF-01102.11-97.A   awarded by the Space  Telescope  Science
  Institute, which  is operated  by AURA  for NASA  under contract NAS
  5-26555.
\end{acknowledgments}

%
% Bibliography
%

\end{document}